\documentstyle[prl,aps,twocolumn]{revtex}

\title{A note on covariant action integrals in three dimensions}
\author{M\'aximo Ba\~nados$^{1,2,}$\thanks{On leave from Departamento
de F\'{\i}sica, Universidad de
Santiago de Chile, Casilla 307, Santiago 2, Chile} and Fernando
M\'endez$^{2,3}$}
\address{\it $^1$Departamento de F\'{\i}sica
      Te\'orica, Facultad de Ciencias, Universidad de Zaragoza,
      Zaragoza 50009, Spain \\
$^2$Centro de Estudios Cient\'{\i}ficos de Santiago, Casilla 16443,
Santiago, Chile \\
$^3$Departamento de F\'{\i}sica, Facultad de Ciencias, Universidad de Chile, Casilla 653, Santiago, Chile.
}

\begin{document}

\maketitle
\begin{abstract} 
We compute -in the saddle point approximation- the partition function
for the 2+1 black hole using the Gibbons-Hawking approach. Some issues
concerning the definition of thermodynamical ensembles are clarified.
It is pointed out that the right action in covariant form is exactly
equal to the Chern-Simons action with no added boundary terms. This
action is finite, yields the right canonical free energy and has an
extremum when the temperature and angular velocity are fixed. The correspondence with a 1+1 $CFT$ is indicated.\\

\end{abstract}

\section{Introduction}

In the standard approach to covariant Euclidean quantum gravity
\cite{gh}
one considers the functional integral
\begin{equation}
Z[h] = \int Dg\, e^{-I[g,h]}
\label{Z}
\end{equation}
with
\begin{equation}
I[g,h] = -\frac{1}{16\pi G}\left(\int_M \sqrt{g} R + 2\int_{\partial
M} \sqrt{h} K \right)  
\label{I0}
\end{equation}
and $h_{ij}$ is the 3-metric induced on $\partial M$. The boundary
term is added to the action to ensure that $I$ has an extremum when
$h$ is fixed; the partition function thus depends on the 3-metric $h$,
and the measure $Dg$ denotes the sum, modulo diffeomorphisms, over all
metrics with $h_{ij}$ fixed. Despite the unrenormalizible
divergences appearing in the perturbation expansion of this functional
integral, the formal expression (\ref{Z}) has provided interesting
insights to quantum gravity \cite{gh}.  The idea is that although
(\ref{Z}) cannot be computed for higher loops, its saddle point
approximation around some classical solutions does give reasonable,
and indeed correct, results. The first example of a successful
evaluation of (\ref{Z}) was performed by Gibbons and Hawking
\cite{gh} who considered an Euclidean Schwarzschild black
hole of mass $M$ with metric
\begin{equation}
ds^2 = \left( 1 - \frac{2M}{r} \right)dt^2 + \left( 1 - \frac{2M}{r}
\right)^{-1}dr^2 + r^2 d\Omega^2. 
\label{sch}
\end{equation}
An important property of this metric is that, in order to avoid a conical singularity in the plane $r/t$, the coordinate $t$ must
be periodically identified with period
\begin{equation}
\beta=8\pi M.
\label{beta}
\end{equation}
In other words, demanding (\ref{sch}) to be a smooth solution of Einstein
equations fixes the period of Euclidean time $t$ as in (\ref{beta}).
One expects $\beta$ to be related to the inverse temperature of the
black hole. Indeed, the value (properly regularized, see below) of $Z$
in the saddle point (\ref{sch}) is 
\begin{equation}
Z[\beta] \sim e^{-\beta^2/16\pi}.
\label{Zsch}
\end{equation}
The thermodynamical formula for the average energy $M = -\partial\log
Z/\partial \beta$ reproduces (\ref{beta}) and confirms that $\beta$ is
the inverse temperature of the black hole. This result has two
important consequences: (i) it shows that the formal expression
(\ref{Z}) does contain relevant and correct information about quantum
gravity, (ii) the black hole thermal properties are present in a pure
quantum theory of gravity, and not only in the interaction of a
classical background metric with quantized fields.   

In the evaluation of $I$ on the black hole manifold (\ref{sch}) there
is a divergence which is canceled by subtracting the value of $K$ for
$M=0$ \cite{gh}.  The goal of this note is to clarify some issues
concerning the regularization procedure and it consequences in the
definition of ensembles, in three dimensions. In particular, we shall
exhibit a covariant form for the regularized action which does not
seems to exist in four dimensions. Other approaches for this problem
have been studied in  \cite{Brown-Creighton-Mann}. A recent review
about three dimensional black hole statistical mechanics and
thermodynamics can be found in \cite{Carlip98}.

The evaluation of expressions like (\ref{Z}) on a saddle point has
become crucial in the recently conjectured $adS/CFT$ correspondence
\cite{Maldacena}. This conjecture has been reinterpreted in
\cite{Gubser-,Witten98} in terms of generating functions and in this
formulation the evaluation of the action (\ref{I0}) on classical
solutions is a key element. Furthermore, 
it has been pointed out \cite{Witten98} that the infinities
appearing in the evaluation of $I$ represent expected anomalies from
the point of view of the boundary conformal field theory. It is thus
an important point to understand clearly where the infinities come
from. The particular case of three dimensional supergravity is also
important because the related conformal field theory is 2 dimensional.     

\section{Ensembles and the Hamiltonian approach }

Let us study in some detail the evaluation of $I$ on the
saddle point (\ref{sch}). First, note that we have expressed the value
of $Z$ in terms of $\beta$ rather than $M$ which is the `natural'
parameter of the Schwarzschild metric.  The relevant question here is,
why have we assumed that the value of $Z$ on the saddle point
(\ref{sch}) is associated to a canonical ensemble? Of course one could
argue that it has the right value for a canonical partition function.
There is, however, a better reason: In the action principle
(\ref{I0}), the metrics (\ref{sch}) provide an extremum only if
$\beta$, the Euclidean time period, is fixed.  Then, by
definition, the ensemble is canonical\footnote{We say that the
ensemble is canonical when the temperature is fixed and microcanonical
when the energy is fixed. We shall follow here the most common
approach assuming asymptotic flatness or an anti-de Sitter behavior.
In both cases, the concept energy --as a conserved quantity associated
to asymptotic time translations--  can be defined. We also define the
inverse temperature $\beta$ as the period of the Euclidean time
coordinate used to define the energy. In asymptotically flat
spacetimes this coordinate coincides with proper time, but this is not
the case for asymptotically anti-de Sitter spaces.} and it is correct
to express $Z$ in terms of $\beta$.  The consistency between the
variational principle and the actual value of $Z$ is automatic if one
works in the Hamiltonian formalism (see below) and probably this is
why most people do not worry about this issue. However, in the
Lagrangian formulation there is a non-trivial regularization in the
calculation of $I$ (the boundary term in (\ref{I0}) has to be
regularized by subtracting its value for the zero mass black hole) and 
therefore this consistency cannot be taken for granted. 

In the Hamiltonian approach to quantum gravity the appropriate action
principle for metrics approaching (\ref{sch}) is
\begin{eqnarray}
I &=& \int (\pi^{ij} \partial_t g_{ij} + N {\cal H} + N^i {\cal H}_i
)d^3xdt  \nonumber\\ && +  \beta M - \frac{A(r_+)}{4G}
\label{IH}
\end{eqnarray}
where $\pi^{ij}$ is the momentum conjugate to the spatial metric
$g_{ij}$,  ${\cal H}$ and ${\cal H}_i$ are the generators of normal
and tangential deformations respectively and $N,N^i$ their associated
Lagrange multipliers. The boundary term $\beta M$ is added to cancel a
boundary term (at infinity) equal to $-\beta \delta M$ coming from an
integration by parts in the bulk. Thus, the variation of (\ref{IH}) is
well defined at infinity provided $\beta$ is fixed which means that
the relevant ensemble is the canonical one.    

The second term is located at the horizon (see, for example,
\cite{Brown-,Wald93,BTZ4}). A boundary term at the horizon can be
expected in the Hamiltonian formalism because the time coordinate is
degenerate there. Indeed, the Killing vector field $\partial_t$ has a
fixed point at the horizon and thus the foliation becomes degenerate
at that point. One way to fix this problem \cite{BTZ4} is to separate
the action into two pieces. One can use the covariant action
(\ref{I0}) on a small disk of radius $r=\epsilon$ around the horizon
and, for $r>\epsilon$, we use
the Hamiltonian action. As we let $\epsilon$ go to zero, one
discovers that the value of the covariant action approaches the value
$A/4G$, rather than zero. This can be understood in terms of the
Gauss-Bonnet theorem\cite{BTZ4}. In the limit $\epsilon\rightarrow 0$,
the covariant action becomes proportional to the Euler number for a
disk times the area of the horizon ($A$), which is different from
zero. Since the Euler number is a topological invariant, it does not
change as we let $\epsilon\rightarrow 0$. In the limit one finds a 
contribution to the action equal to $A/4G$ \cite{BTZ4}.  Now we study
the variation of the action (\ref{IH}) at the horizon. 
The boundary term coming from the variation of the bulk is exactly
equal to $\delta (A/4G)$ and is canceled by the boundary term. The
action principle is thus well defined.     
 
To evaluate the action (\ref{IH}) on the saddle point (\ref{sch}) we
note that since the metric is static and it satisfies the constraints
one is left only with the boundary terms. They clearly provide the
right form for the free energy of the black hole.  Expresing $M$ and
$A=4\pi r_+^2$ in terms of $\beta$ one finds $I = \beta^2/16\pi$, as
expected. The Hamiltonian approach to quantum black holes has been
applied to a large number of situations, even with Lovelock
Lagrangians \cite{Jacobson-,BTZ4}.  It is a generic feature of this
approach that the entropy comes from a boundary term at the horizon
while infinity contributes with the term  $-\beta M$, plus other
potentials if more charges are present.  

At this point we can make a precise comparison between the Lagrangian
and Hamiltonian approaches. In the Lagrangian formalism, all boundary
terms are located at infinity and it is not necessary to give any sort
of boundary conditions at the horizon. However, to define the value of
the action on the saddle point (\ref{sch}), a non trivial
regularization is needed \cite{gh}. In the Hamiltonian approach, there
are both boundary terms at infinity and at the horizon, but the action
is finite on the saddle point.  Finally, the definition of ensembles
and its consistency with the value on the saddle point is direct in
the Hamiltonian formulation. In the Lagrangian formalism this needs to
be checked explicitly.

\section{The three dimensional case. Lagrangian approach}

In this section we shall study the quantum 2+1 black hole in the
Lagrangian formalism and show explicitly the consistency between the 
value of the regularized action and the ensemble. In the next section,
we shall introduce a covariant form for the regularized action. The
thermodynamics of anti-de Sitter space in 3+1 dimensions was first
studied in \cite{Hawking-Page}.  

The Euclidean three dimensional black hole solution with angular
momentum is \cite{BTZ}
\begin{equation}
ds^2 = \beta ^2 N^2 dt^2 + N^{-2} dr^2 + r^2 (\beta N^{\varphi} dt +
d\varphi)^2
\label{3}
\end{equation}   
with
\begin{eqnarray}
N^2     & = & \frac{r^2}{\ell ^2} - M - \frac{J^2}{4 r^2} \\
N^\varphi & = & \frac{J}{2 r^2} + \Omega
\end{eqnarray}
The ranges are $0\leq t<1$ (note that we have introduced a parameter
$\beta$ in the metric equal to the Euclidean time period), $r_+ \leq
r<\infty$, where $r_+$ is the larger solution of $N^2=0$, and $0\leq
\varphi <2\pi$.  The parameters $M$ and $J$ are the mass and angular
momentum respectively, while $\Omega$ the Euclidean angular velocity.  

We start with the Euclidean action for 2+1 gravity with a negative
cosmological constant
\begin{equation} 
I_{0} = -\frac{1}{2 \pi} \int \left( R + \frac{2}{\ell^2} \right)
\sqrt{g} d^3x,
\label{Io}
\end{equation}
where we have chosen $8G=1$.  (One can easily check that, as in four
dimensions \cite{Hawking-Page}, the regularized boundary term $2\int
(K-K_0)$ is zero for anti-de Sitter spacetimes.)

As a first step we evaluate this action on the saddle point (\ref{3})
and find
\begin{equation}
I_0 = \frac{ 2 \beta}{\ell ^2} (r_{\infty} ^2 - r_{+} ^2), 
\label{in}
\end{equation} 
where $r_{\infty}$ is the value of $r$ at infinity. Of course, we need
to regularize this expression. Note that the calculation of $I_0$ is
direct because the black hole has constant negative curvature equal
to $R=-6/l^2$. 

A natural regularization for (\ref{in}) is to subtract the
value of $I_0$ on the vacuum black hole obtained from (\ref{3}) by
setting $M=J=\Omega=0$. In this limit the metric becomes
\begin{equation}
ds^2 = \beta^2 r^2 dt^2 + \frac{dr^2}{r^2} + r^2 d\varphi^2.
\label{vac}
\end{equation}
with the ranges $0\leq t<1$, $0<r<\infty$ and $0\leq \varphi<2\pi$. 
Note that $\beta$ is no longer constrained to any particular value.
However, in the subtraction procedure \cite{gh}, we take
the metric (\ref{vac}) with $\beta$ equal to the value of the non-zero
mass black hole. The value of (\ref{Io}) on the vacuum metric is
$-2\beta r^2_\infty/l^2$ and thus the regularized action becomes,
\begin{equation}
I_{reg} = -\frac{1}{2 \pi} \int \left( R + \frac{2}{\ell^2} \right)
\sqrt{g} d^3x - \frac{2\beta r_\infty^2}{l^2}
\label{Ireg}
\end{equation}
and its value on the black hole is $I_{reg}= -2 \beta r_+ ^2/\ell^2$. 

Now we need to determine whether this value for $I$ corresponds to a
canonical, microcanonical or a mixed ensemble.  In other words, we
don't know whether to express the value of $I$ in terms of the
intensive or extensive parameters. This information is provided by the
variations of the action. 

The general variation of (\ref{Io}) has the form
\begin{eqnarray}
\delta I_0  & = &   \int_{B}( eom )\sqrt{g} d^3 x  \nonumber \\ 
            &   & - \frac{1}{2  \pi} \int_{ \partial B} \sqrt {g} 
g_{ \sigma \rho} ( n_{ \alpha} g^{ \mu \nu}  \Gamma ^{ \rho}_{ \mu
\nu} -n^{ \mu}\Gamma ^{ \rho}_{ \alpha \mu} ) \delta g^{ \alpha
\sigma} d^2
x   \nonumber \\
            &   &  - \frac{1}{2  \pi} \int_{ \partial B} \sqrt {g}
            n_\lambda
( g^{\mu \lambda} g^{\nu \sigma} -  g^{\mu \nu} g^{\lambda \sigma} )
\delta g_{\mu \nu , \sigma} d^2 x.
\end{eqnarray}
Since we are interested in the variations of the regularized action
(\ref{Ireg}), we add the subtraction term $-2\beta r_\infty^2/l^2$ and 
evaluate all boundary terms in the black hole metric (\ref{3}).
Keeping all variations $\delta M,\delta J,\delta \beta$ and $\delta
\Omega$, one finds for the variation of (\ref{Ireg})  
\begin{equation}
\delta I_{reg} = \mbox{[e.o.m.]} \, - J \delta
(\beta \Omega) - \beta \delta M 
\label{var0}
\end{equation}
where [e.o.m.] represents the term proportional to the equations of
motion.  
We see that the variation of (\ref{Ireg}) is finite. However,
(\ref{Ireg}) does not represent any of the standard ensembles because
it has well defined variations under $M$ and $\beta\Omega$ fixed. 

This is easily fixed. Consider 
\begin{equation}
I_c = I_{reg} + \beta M.
\label{c}
\end{equation}
This action has well defined variations for $\beta$ and $\Omega$
fixed, and
its value on the saddle point is $\beta (M + \Omega J) - A/4G$, as
expected for a canonical ensemble (we have restored $G$).   As usual,
the microcanonical action can be obtained by subtracting $\beta (M +
\Omega J)$.

\section{The 2+1 black hole canonical covariant action} 

We have found in the last section a Lagrangian action, $I_c$,  
whose value on the saddle point is consistent with its variations. The
action $I_c$ is, however, not covariant. We shall now show that $I_c$
is exactly equal to the Chern-Simons action for three dimensional
gravity, with no added boundary terms.  

First, let us look at the relation between the extrinsic curvature and
the boundary terms that we have calculated. It is direct to check that
\begin{equation}
\frac{1}{2 \pi} \int_{\partial B} K \sqrt {h} = 
\beta \left(- M + \frac{2 r^2_\infty}{\ell^2} \right).
\end{equation}
The right hand side contains exactly the two terms that we have added
to (\ref{Io}) in order to make it finite, see (\ref{Ireg}), and to
have well defined variations on the canonical ensemble, see (\ref{c}).

Thus, we arrive at the result that we can write a
covariant formula for the canonical regularized action, 
\begin{equation}
I_c = -\frac{1}{2\pi} \left[  \int \left( R+ \frac{2}{l^2} \right) +
\int_{r=\infty} \sqrt{h} K \right] 
\label{Ifin}
\end{equation}
where the boundary term provides both, the regularization of the bulk
(with the right value on the saddle point) as well as the right
variation.  The action (\ref{Ifin}) is finite, it has well defined
variations for $\beta$ and $\Omega$ fixed, and its saddle point value
is consistent with the canonical ensemble.  In this sense (\ref{Ifin})
is as good as the Hamiltonian action. However, it should be noted that
(\ref{Ifin}) does not have any added boundary terms at the horizon. 

A quick comparison between (\ref{Ifin}) and (\ref{I0}) reveals a
disturbing missing factor of 2 in the boundary term. This means that
the action (\ref{Ifin}) is not appropriate to fix the metric at the
boundary but rather, a combination of the metric and its derivatives.
This is not too surprising. It only means that there isn't a one to
one correspondence between fixing the metric or its derivatives with
fixing the intensive or extensive parameters.    

The surprising fact is that there is a direct geometrical interpretation for the fields which are fixed in (\ref{Ifin}). They correspond to the $SL(2,C)$ gauge field arising in the Chern-Simons formulation of Euclidean three dimensional $adS$ gravity.  It is known that in three dimensions Euclidean gravity  with a negative cosmological constant can be recast as a Chern-Simons theory \cite{Achucarro-,Witten88} for the group $SL(2,C)$. The Einstein-Hilbert and Chern-Simons actions differ by a boundary term which is exactly the one appearing in (\ref{Ifin}).

This can be seen as follows. Let us consider the Einstein-Hilbert
action in the first order formulation
\begin{equation}
\int \sqrt{g} \left(R + \frac{2}{l^2} \right) = \int \epsilon_{abc}
\left(R^{ab} + \frac{e^a \wedge e^b}{3l^2} \right) \wedge e^c. 
\label{Ie}
\end{equation}
In this action, the triad $e^a$ appears with no derivatives. Then,
when computing the variations one picks up a boundary term
$\epsilon_{abc} e^a \delta w^{bc}$ which means that we need to fix the
spin connection, or in metric language, a function of the metric and
its derivatives.  If one wants to have an action in which only the
metric needs to be fixed, one can  make an integral by parts and pass
the derivative to the triad. The necessary boundary term is $\int
\epsilon_{abc} w^{ab} \wedge e^c$, and it is equal to $2 \int \sqrt{h}
K$, as expected.  

However, a more interesting modification of (\ref{Ie}) occurs if one
removes one half of the above boundary term: $(1/2) \int
\epsilon_{abc} w^{ab} \wedge e^c$ (which is the term that has been
added in (\ref{Ifin})). In this case, both the triad and the spin
connection have derivatives, and the action reduces exactly to the
well known Chern-Simons form
\begin{equation}
I_c = iI[A]- iI[\bar A]. 
\label{ICS}
\end{equation}
with $A^a =(1/2) \epsilon^a_{\ bc} w^{bc} + i e^a/l$, $\bar A =(1/2)
\epsilon^a_{\ bc} w^{bc} - i e^a/l $, $k=-l/4G$ and 
\begin{equation}
I[A] = \frac{k}{4\pi} \int Tr (AdA + \frac{2}{3} A^3 ).
\end{equation}
This action has well defined variations if one fixes the values of one
component of the gauge field at the boundary. In the Euclidean black
hole topology -a solid torus- the most convenient boundary conditions
are simply \cite{BBO} $A_{\bar z}=0$ and $\bar A_z=0$. Indeed, the
black hole field satisfies these conditions and they yield a Kac-Moody
symmetry lying at the boundary.  

It is instructive to check that the on-shell value of (\ref{ICS}) does
yield the right canonical free energy.  Since the black hole has zero
curvature $dA + AA=0$, the integral to be computed is, as in instantons
calculations, $-(1/3) \int A^3$.  The formulae for the black hole
gauge field $A$ in Schwarzschild coordinates can be found, for
example, in \cite{BBO}. However, if one naively replaces $A$ in the
above integral it yields zero. The reason is that the  Schwarzschild
coordinates are singular at the horizon.  Another coordinate system
are the Kruskal coordinates which are regular at the horizon but
singular at infinity. To compute the value of $I$ one then needs to
split the integral into two patches. The calculation is further
complicated by the fact that the Chern-Simons action is not gauge
invariant -- it transforms by a boundary term. Thus, one has to make
sure to use the same gauge (one consistent with the boundary
conditions) in the two patches. 

Fortunately, there is a quicker way to arrive at the right result by
using  angular quantization. Since the black hole field does not
depend on $\varphi$, and the vector field $\partial_\varphi$ is well
defined in the whole solid torus, it is useful to use this coordinate
as time, and make a Hamiltonian decomposition for the Chern-Simons
action. 
One obtains (taking one copy)
\begin{equation}
I = \frac{k}{4\pi} \left( \int_M \epsilon^{ij}(- A_i \partial_\varphi
A_i + A_\varphi F_{ij}) 
+ \int_{T^2} A_0  A_\varphi \right) 
\label{IHCS}
\end{equation}
with $x^i=\{t,r\}$. The only boundary of the solid torus is at the
torus at infinity \footnote{Note that, just as in the metric
formulation, if one does a time Hamiltonian decomposition, there is
also a boundary term at the horizon because the vector field
$\partial_t$ is not well defined there. The resulting action has been
considered in \cite{BBO}.}.  The on-shell value of (\ref{IHCS}) is
easy to compute  because the black hole field satisfies the constraint
$F_{ij}=0$ and it does not depend on $\varphi$. The only contribution
is then the boundary term which can be computed using Schwarzschild
coordinates. Collecting both copies of the Chern-Simons action and
using the formulae for $A$ given in \cite{BBO} one finds $I_c = -\beta
(M + \Omega J)$, which thanks to the three dimensional Smarr formula
\begin{equation}
\beta (M + \Omega J) = - \beta (M + \Omega J) + S
\label{Smarr}
\end{equation}
yields the right free energy. Note that in time quantization
\cite{BBO} the value of the action is equal to the right hand side of
(\ref{Smarr}) with the term $\beta (M + \Omega J)$ coming from the
boundary term at infinity, while $S$ comes from the horizon. In
angular quantization there is a single boundary term at infinity whose
value is the left hand side of (\ref{Smarr}). 

We have shown that the right action for studying the properties of a
quantum black hole is the Chern-Simons action without any added
boundary terms. It would be very interesting to see whether this
property is carried over to higher odd dimensional spacetimes where a
Chern-Simons formulation for gravity \cite{Chamseddine} as well as
black hole solutions \cite{BTZ3} exist. 

Finally, let us comment on the meaning of our results in the light of
the $adS/CFT$ correspondence \cite{Maldacena}.  In the form displayed in
\cite{Gubser-,Witten98}, the partition function of the three dimensional
supergravity theory, as a function of the boundary conformal structure 
$h$, should be equal to the partition function of the 1+1 $CFT$,
\begin{equation}
Z_{sugra} (h)  = Z_{CFT} (h). 
\label{ZZ}
\end{equation}
If the 1+1 $CFT$ theory has zero central charge, then $Z_{CFT}$ depends only on the conformal structure. If $c\neq 0$, then $Z_{CFT}$ depends also on the conformal factor.  In the conformal
gauge, $Z_{CFT}(h)$ gives rise to a Liouville action for the conformal factor  with a coupling constant equal to the central charge.  Now
we look at the left hand side of (\ref{ZZ}). In view of the results
obtained in \cite{CHvD}, which are derived using the same boundary
conditions that we have used (see also \cite{BBCHO} for a
supersymmetric extension), we find that, indeed, $Z_{sugra}$ is given
by a Liouville action.  The central charge of the supergravity theory
is known to be $c=3l/2G$ \cite{BH}, and that fixes the central charge
of the corresponding $CFT$.  This provides another example where a
classical property of the supergravity theory (the central charge in
this case) has a quantum origin in the corresponding $CFT$. An
intriguing consequence of this correspondence is the fact that the
number of states in the $CFT$ with this central charge give rise exactly
to the three dimensional black hole entropy \cite{Strominger}.  In
this context, our formulation in terms of a single boundary at
infinity may shed some light on this interesting subject. 

Recently \cite{K}, the three dimensional $adS$ central charge has been derived using the regularization method outlined in \cite{Witten98}. In  this procedure, the central charge is associated with the infinities appearing in the classical supergravity action since the corresponding counterterms break conformal invariance.  In our approach, conformal invariance is broken by the constraints on the $SL(2,\Re)$ currents ($A^1=1$ and $A^3=0$) which transmute the affine $SL(2,\Re)$ 
algebra into a Virasoro algebra with $c=6k$ \cite{Polyakov}.\\

We would like to thank T. Brotz and C. Mart\'{\i}nez for a critical reading of the manuscript, and  S. Carlip, M. Henneaux and M. Ortiz for useful discussions.

\end{document}